\documentclass[a4paper,fleqn,usenatbib]{mnras}

\usepackage{mathptmx}

\usepackage[T1]{fontenc}
\usepackage{ae,aecompl}

\usepackage{graphicx}	
\usepackage{amsmath}	
\usepackage{amssymb}	

\newcommand{\Msun}{\,\mathrm{M_{\sun}}}

\usepackage[dvipsnames]{xcolor}

\title[Pre-processing \& dynamical dependence]{Evidence of
  pre-processing and a dependence on dynamical state for low-mass
  satellite galaxies}

\author[I.D. Roberts \& L.C. Parker]{
Ian D. Roberts,\thanks{E-mail: roberid@mcmaster.ca}
Laura C. Parker
\\
Department of Physics and Astronomy, McMaster University, Hamilton ON
L8S 4M1, Canada
}

\date{Accepted XXX. Received YYY; in original form ZZZ}

\pubyear{2016}

\begin{document}
\label{firstpage}
\pagerange{\pageref{firstpage}--\pageref{lastpage}}
\maketitle

\begin{abstract}
We study the dependence of satellite star formation rate and
morphology on group dynamics for a sample of SDSS groups.  We classify
the group dynamical state and study satellite properties for populations of
galaxies at small and large group-centric radii.  For galaxies at large radii
we find no differences in the star-forming or
disc fraction for those in Gaussian groups compared to
those in non-Gaussian groups. By comparing the
star-forming and disc fractions of infalling galaxies to field
galaxies we find evidence for the pre-processing of both star
formation rate and morphology.  The strength of pre-processing
increases with halo mass and is highest for low-mass galaxies
infalling onto high-mass haloes. We show that the star formation rate of galaxies at small radii correlates with group
dynamical state, with galaxies in non-Gaussian groups showing
enhanced star-forming fractions compared to galaxies in
Gaussian groups.  Similar correlations are not seen for the disc fractions of galaxies at small radii.  This seems to suggest that either the
mechanisms driving star formation quenching at small halo-centric radii are more efficient in
dynamically relaxed groups, or that non-Gaussian groups have assembled more
recently and therefore satellites of the groups will have been exposed
to these transforming mechanisms for less time.
\end{abstract}




\section{Introduction}
\label{sec:introduction}

In the first half of the twentieth century it was beginning to be
realized that populations of high-mass clusters were predominantly
made up of early-type galaxies, with \citet{hubble1931} stating that,
``The predominance of early types is a conspicuous feature of clusters
in general''.  Many subsequent
observational studies have cemented the now
familiar environmental dependence of galaxy properties
\citep[e.g.][]{butcher1978, dressler1980, postman1984, dressler1999,
  blanton2005, wetzel2012}.  Namely, galaxies in clusters tend
to be red in colour with low star formation
rates and early-type morphologies.  On the other hand the
low-density field is preferentially populated by blue, star
forming, spiral galaxies.  A third environment, galaxy groups, are the
most common environment in
the local Universe \citep{geller1983, eke2005} and also represent an
intermediate-mass regime in which significant populations of both
star-forming spirals and passive ellipticals are observed
\citep[e.g.][]{wilman2005, mcgee2011}.
\par
Not only do galaxy properties correlate with the type of haloes in
which they reside, but also with distance from the halo centre.
In particular, galaxies at large radii show enhanced star
formation and are more likely to have spiral morphologies compared to
galaxies near the centre of the halo \citep{whitmore1993, goto2003,
  postman2005, rasmussen2012, wetzel2012, fasano2015, haines2015}.
Therefore, in order to probe the environmentally driven aspects of
galaxy evolution it is crucial to account for both the dependence on
the host halo environment as well as the radial position within the
group or cluster.
\par
The aforementioned environmental dependences are strongest for
low-mass galaxies and it appears that properties of high-mass galaxies
are less dependent on environment \citep{haines2006, bamford2009}.  For
high-mass galaxies, quenching is thought to be driven by internal, secular
processes such as feedback from AGN \citep[e.g.][]{schawinski2009}.  This
dichotomy between high and low mass galaxies is 
presented in \citet{peng2010} where it is argued that in
the local Universe galaxies below $\sim\!10^{10.5}\Msun$ are
environmentally quenched as satellite galaxies and galaxies above that
mass are primarily quenched by internal processes (so-called ``mass
quenching'').
\par
While it appears that the majority of low-mass galaxies are primarily quenched
as satellites, there are still open questions regarding the details of
the process(es) involved.  One such question is which are the dominant
mechanism(s)
responsible for suppressing star formation in satellite galaxies?
Galaxy harassment \citep[e.g.][]{moore1996}, mergers \citep[e.g.][]{mihos1994},
starvation \citep[e.g.][]{kawata2008}, and ram-pressure stripping
\citep[e.g.][]{gunn1972} have all been invoked, however no concensus
exists on their relative importance in different environments.
Additionally, while all of these mechanisms are capable of
quenching galaxies (either through inducing rapid star formation and
thus quickly using up cold gas reserves, or
the stripping of gas), not all would have a strong effect on galaxy
morphology.  Recently, starvation and/or ram-pressure stripping are
often favoured
as satellite quenching mechanisms \citep{muzzin2014, peng2015, fillingham2015,
  weisz2015, wetzel2015} but it is not clear that either
would strongly impact morphology, therefore in order to explain the
observed correlation between galaxy star formation and morphology it
seems that an additional process to efficiently drive morphological
transformations is perhaps required \citep[e.g.][]{christlein2004}.
\par
Also of importance is determining
the characteristic haloes in which most satellite galaxies are
quenched and experience morphological transformations.  Do galaxies remain actively forming
stars with late-type morphologies until passing the virial radius of high-mass clusters, or
are they transformed in smaller groups prior
to or during cluster infall (known as ``pre-processing'')
\citep[e.g.][]{fujita2004}.  Pre-processing is
often invoked to explain observational results such as passive and
red fractions at large cluster-centric radii which are enhanced
significantly relative to the field \citep{lu2012, wetzel2012,
  bahe2013, haines2015, just2015}, as well as the prevalence of S0
galaxies in large clusters \citep{kodama2001, helsdon2003, moran2007,
  wilman2009}.  Studies
have also found evidence for pre-processing by measuring the fraction
of galaxies which are part of a group subhalo during infall onto a cluster,
both using simulations \citep{mcgee2009, delucia2012, bahe2013} and
observations \citep{dressler2013, hou2014}.
\par
This pre-processing and recent infall of galaxies can imprint
itself on the dynamical profile of a group or cluster.  For a
dynamically relaxed group it is expected that the projected velocity profile of member
galaxies will resemble a Gaussian distribution whereas groups which
are dynamically young and unrelaxed tend to display velocity
profiles which are less Gaussian in nature \citep[e.g.][]{yahil1977, bird1993, martinez2012, ribeiro2013}.  The degree to which
galaxy properties correlate with the dynamical state of their host
groups is still an open question \citep[e.g.][]{biviano2002,
  ribeiro2013b}, though it may be expected that such correlations exist.  For example, dynamically complex groups are preferentially X-ray underluminous \citep{popesso2007, roberts2016} which indicates an underdense intra-group medium.  Considering that many quenching mechanisms operate through interactions with the intra-group medium it may be expected that such mechanisms will be less efficient in non-Gaussian groups.  Furthermore, if non-Gaussian groups represent younger systems then galaxy properties could be affected (compared to Gaussian systems) simply due to galaxies being exposed to a dense environment for less time.  Previous work has suggested that 
galaxies in relaxed groups tend to be redder than galaxies in
unrelaxed groups \citep{ribeiro2010, carollo2013, ribeiro2013}.
However less work has been done studying the dynamical dependences of
star-formation and morphology directly.  One example is the work of
\citet{hou2013} who find no detectable difference between the
quiescent fractions
of galaxies in Gaussian versus non-Gaussian groups as a function of redshift.
\par
Previously we have shown that the star formation and morphology of
low-mass galaxies depends not only on stellar and halo mass but also
on the X-ray luminosity of the host group \citep{roberts2016}.  Here
we investigate the dependence of star-forming and 
morphological properties of galaxies on group dynamical state.  In
particular, we study these properties within different radial regions
of the halo to explore whether galaxy properties correlate with group dynamical state and whether any correlations show radial dependence.
\par
The outline of this paper is as follows.  In Section~\ref{sec:data} we
describe the sample of galaxies in groups as well as our field sample. In
Section~\ref{sec:r_high} we analyze the dependence of galaxy star
formation and morphology on dynamics for galaxies at large radii.  In
Section~\ref{sec:r_low} we do the same for galaxies in the inner regions of the halo.  We
discuss our results in Section~\ref{sec:discussion} and summarize in
Section~\ref{sec:summary}.
\par
In this paper we assume a flat $\Lambda$ cold dark matter cosmology
with $\Omega_\mathrm{M} = 0.3$, $\Omega_\mathrm{\Lambda} = 0.7$, and
$H_0 = 70\,\mathrm{km}\,\mathrm{s^{-1}}\,\mathrm{Mpc^{-1}}$.


\section{Data}
\label{sec:data}

\subsection{Group sample}
\label{sec:data_group}

For this work we employ the group catalogue of \citet{yang2007}, which
is constructed by applying the halo-based galaxy group finder from
\citet{yang2005, yang2007} to the New York University Value-Added
Galaxy Catalogue (NYU-VAGC; \citealt{blanton2005}).  The NYU-VAGC is a
low redshift galaxy catalogue consisting of
$\sim\!700\,000$ galaxies in the Sloan Digital Sky Survey
Data Release
7 (SDSS-DR7; \citealt{abazajian2009}).  We will briefly describe the
halo-based group finding algorithm used to generate the Yang group catalogue,
however for a more complete description please see \citet{yang2005} and \citet{yang2007}.
\par
First, the centres of potential groups are identified.  Galaxies are
initially assigned to groups using a traditional
``friends-of-friends'' (FOF) algorithm \citep[e.g.][]{huchra1982} with
very small linking lengths.  The luminosity-weighted centres of
FOF groups with at least two members are then taken as the centres of
potential groups and all galaxies not yet associated with a FOF group
are treated as tentative centres for potential groups.  A
characteristic luminosity, $L_{19.5}$, defined as the combined
luminosity of all group members with $^{0.1}M_r - 5\log h \le -19.5$,
is calculated for each tentative group and an initial halo mass is
assigned using an assumption for the group mass-to-light ratio,
$M_H/L_{19.5}$.  Utilizing this tentative group halo mass, velocity
dispersions and a virial radius are calculated for each group.  Next,
galaxies are assigned to groups under the assumption that the
distribution of galaxies in phase space follows that of dark matter
particles -- the distribution of dark matter particles is assumed to
follow a spherical NFW profile \citep{navarro1997}.  Using the new
group memberships, group centres are recalculated and the procedure is
iterated until group memberships no longer change.
\par
We take group halo masses, $M_H$, from the Yang catalogue calculated
using a characteristic group stellar mass, $M_{\star,\text{grp}}$, and
assuming that there is a one-to-one relation between $M_{\star,\text{grp}}$
and $M_H$.  \citet{yang2007} define $M_{\star,\text{grp}}$ as

\begin{equation}
  M_{\star,\text{grp}} = \frac{1}{g(L_{19.5},\,L_{\text{lim}})} \sum_i
  \frac{M_{\star,i}}{\mathcal{C}_i}
\end{equation}

\noindent
where $M_{\star,i}$ is the stellar mass of the $i$th member galaxy,
$\mathcal{C}_i$ is the completeness of the survey at the position of
that galaxy, and $g(L_{19.5},\,L_{\text{lim}})$ is a correction factor
which accounts for galaxies missed due to the magnitude limit of the
survey.  While we utilize halo masses derived from group stellar mass in this paper, we have run the same analysis using halo masses derived from group luminosity in the Yang catalogue and see no changes in observed trends.  \citet{campbell2015} show that the choice between stellar mass and luminosity as a halo mass predictor can introduce biases in mass estimates.  For example, when group luminosity is assumed to be the primary property determining halo occupation in mock catalogues, halo masses inferred from group stellar mass are systematically larger for haloes with a red central compared to haloes with a blue central \citep{campbell2015}.  For the samples of Gaussian and non-Gaussian groups which are frequently compared in this paper (see Section~\ref{sec:grp_dyn}), we find that the fraction of groups with passive centrals is 94 per cent in both cases, therefore the aforementioned effects should not preferentially bias one sample more than the other.
\par
The Yang catalogue contains both haloes which would be broadly classified as
groups ($10^{12} \la M_H \la 10^{14}\Msun$) as well as clusters ($M_H
\ga 10^{14}\Msun$), however for brevity we will refer to all haloes as
groups regardless of halo mass unless otherwise specified.
\par
We calculate group-centric radii for all group members within the
sample using the redshift of the group and the angular separation of the galaxy from the
luminosity-weighted centre of the host halo.  Radii are
normalized by the virial radius, $R_{200}$, of the group which is defined as \citep{yang2007, tinker2008}:

\begin{equation} \label{eq:r200}
  R_{200} = \left[\frac{M_H}{200(4/3)\pi \Omega_{m,0} \rho_{c,0} (1+z)^3}\right]^{1/3}.
\end{equation}

\noindent
For the cosmology assumed in this work, equation~\ref{eq:r200} becomes

\begin{equation}
  R_{200} = 1.13\,h^{-1}\,\mathrm{Mpc}\,\left(\frac{M_H}{10^{14}\,h^{-1}\Msun}\right)^{1/3}\,(1 + z_{\mathrm{group}})^{-1}.
\end{equation}

\noindent
For our group sample, we consider galaxies which have projected group-centric radii within $\mathrm{R_{200}}$.
\par
To study specific characteristics of galaxies within the group
sample, we match various public SDSS galaxy catalogues to the group
sample.  We utilize galaxy stellar masses and k-corrected absolute magnitudes given in the NYU-VAGC, which
are obtained through fits to galaxy spectra and broadband photometric
measurements following the procedure of \citet{blanton2007}.
\par
For our star formation indicator we use fibre-corrected specific star formation rates
($SSFR = SFR/M_\star$) from the MPA-JHU DR7 catalogue \citep{brinchmann2004}.  These SSFRs are
primarily derived from emission lines, with an exception for galaxies
with no clear emission lines or AGN contamination in which case SSFRs
are based on the 4000 \AA\ break.  SSFRs for galaxies with $\text{S/N}
> 2$ in H$\alpha$ are determined using only the H$\alpha$ line and
SSFRs for galaxies with $\text{S/N} > 3$ in all four BPT lines are
determined using a combination of emission lines.  For this work we define star-forming galaxies to be all galaxies with $\log SSFR \ge -11$,
\citet{wetzel2012} show that in the local Universe the division
between the red sequence and the blue cloud is consistently found at
$\log SSFR \simeq -11$ across a wide range of halo masses.
\par
For our morphology indicator we use a global S\'{e}rsic
index, $n$, taken from the single component S\'{e}rsic fits in
\citet{simard2011}, and define disc galaxies as all galaxies with $n \le 1.5$.  While the distribution of S\'{e}rsic index is
not as clearly bimodal as the SSFR distribution, we find that our
observed trends are insensitive to our exact choice of dividing S\'{e}rsic
index.  We also weight all of the data by
$1/V_\text{max}$ as given in \citet{simard2011} to account for the stellar-mass
incompleteness of the sample.  This does not explicitly account for the fact that completeness is also a function of galaxy colour, with star-forming galaxies being visible at higher redshift than passive galaxies \citep[e.g.][]{taylor2011}.  While the sample will be biased towards detecting star-forming galaxies at high redshift, we do not expect that this bias will affect galaxies in Gaussian groups differently than galaxies non-Gaussian groups.  Furthermore, the fact that we match all samples by redshift (see Section~\ref{sec:match}) should help to ensure that the completeness (as a function of colour) of the different galaxy samples does not vary substantially.  We note that we do not use a stellar mass complete sample in this work due to the fact that the sample size would be significantly reduced, and in particular, the sample of galaxies in non-Gaussian groups would be very small.
\par
For our analysis we consider only satellite galaxies within groups.
Central galaxies are defined as the most-massive galaxy (MMG) within a group
and subsequently removed from the data set.  We note that it has been shown that the most-massive (or brightest) group galaxy does not always correspond to the group central (ie.\ the galaxy closest to the centre of the potential), for example \citet{skibba2011} show that the fraction of galaxies which are brightest but do not reside at the centre of the potential ranges from $\sim \! 25$ per cent for group-mass haloes to $\sim \! 40$ per cent for high-mass clusters.  To gauge any potential influence that removing the MMG has on the results, we repeat the analysis both with no removal of the MMG and also with removing the second most-massive galaxy instead of the MMG.  In both cases, this does not alter the observed trends qualitatively or quantitatively.
\par
To ensure reasonable statistics when classifying the dynamical states of the groups (see Section~\ref{sec:grp_dyn}) we only include groups from the Yang catalogue which have eight or more member galaxies.  In total, this gives an initial group sample of $47\,961$ galaxies in $2\,662$ groups.

\subsection{Infalling and field samples}

We also define samples of ``infalling'' and ``field'' galaxies, for further comparisons.
\par
To populate the infalling sample we take all galaxies in single-member groups from the Yang catalogue which have projected distances from luminosity-weighted group centres between 1 and 3 virial radii, and have line-of-sight (LOS) velocities less than $1.5\sigma$ from the group centroid, where velocity dispersions, $\sigma$, are calculated using Equation 6 from \citet{yang2007}.  We further define a `strict' infalling sample with the same velocity threshold but only containing galaxies between 2 and 3 virial radii. Galaxies which satisfy this criteria for multiple groups, are assigned as infalling onto the group which they are closest to. This results in a sample of $19\,598$ galaxies infalling onto $2\,396$ groups.
\par
Our field sample is defined as all galaxies in single-member groups which are not members of the infalling sample, and are separated from their nearest `bright' neighbour by at least $1\,\mathrm{Mpc}$ in projected distance and $1\,000\,\mathrm{km}\,\mathrm{s^{-1}}$ in LOS velocity, though our results are insensitive to the exact isolation criteria chosen.  We define bright neighbours as all galaxies which are brighter than the survey r-band absolute magnitude limit at $z=0.2$ (our redshift upper limit), which corresponds to $M_{r,\,\mathrm{lim}} = -21.3$.  Without this condition, the strictness of our isolation criteria would vary with redshift.  We also remove any galaxies which are within $1\,\mathrm{Mpc}$ of a survey edge, or are within $1\,000\,\mathrm{km}\,\mathrm{s^{-1}}$ of our maximum redshift to ensure that all galaxies truly satisfy the isolation criteria.  This yields a field sample consisting of $352\,262$ galaxies.
\par
Stellar masses, absolute magnitudes, SSFRs, and S{\'e}rsic indices for the infall and field sample are obtained from the same sources discussed in Section~\ref{sec:data_group}.

\subsection{Group dynamics}
\label{sec:grp_dyn}

\begin{figure}
  \centering
  \includegraphics[width=\columnwidth]{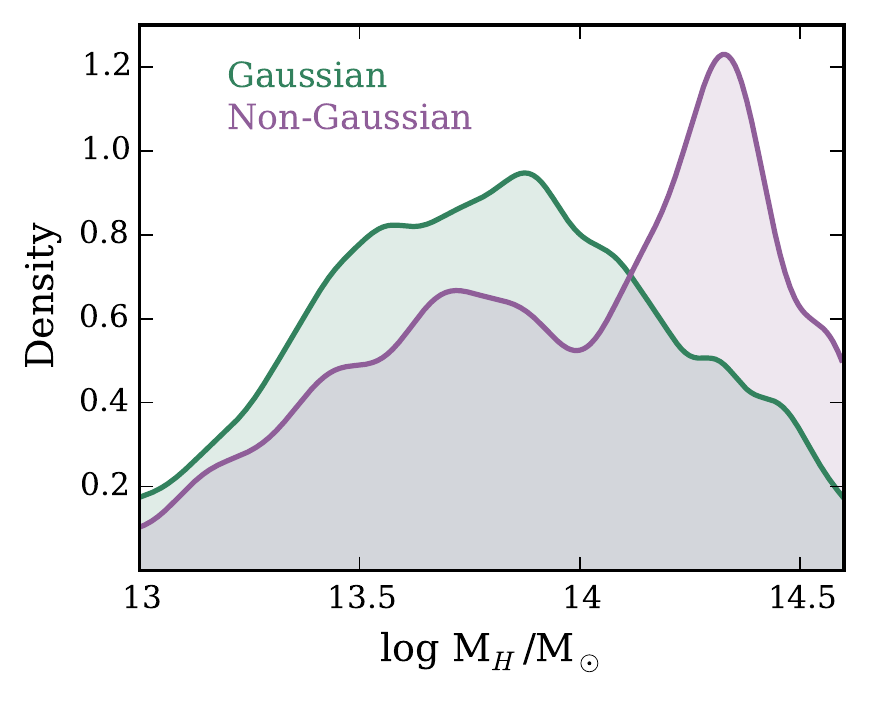}
  \caption{Halo mass distributions, smoothed using a Gaussian kernel, for galaxies in the
    unmatched G and NG samples.}
  \label{fig:mhdist_um}
\end{figure}

To classify the dynamical state of the haloes in the data set we use a
combination of two statistical tests, the Anderson-Darling (AD)
normality test (\citealt{anderson1952}; see \citealt{hou2009, hou2013}
for an astronomical application) and the Dip test
(\citealt{hartigan1985}; see \citealt{ribeiro2013} for an astronomical
application).
\par
The AD test is a non-parametric test of normality based
upon the comparison between the cumulative distribution function (CDF) of a
measured data sample and the CDF of a Gaussian distribution.  Under
the assumption that the data is in fact normally distributed, the AD
test determines the probability ($p$) that the difference between
the CDFs of the data and a normal distribution equals or exceeds the
observed difference.  We apply the AD test to the velocity
distributions of the member galaxies of each group in the sample,
thereby broadly classifying the dynamical state of each halo.  Our
first criteria in classifying a group as Gaussian (G) is that the p-value given
by the AD test be greater than or equal to 0.05.
\par
Our second criteria required for a
group to be classified as G is that its velocity distribution be
unimodal.  Ideally standard normality tests would detect all instances of
multimodality, however this is not always the case.  In particular,
multimodality in distributions with modes at small seperations
can be missed by standard statistical techniques \citep{ashman1994}. To
gauge the modality of the velocity distribution of a given group we
use the Dip test.  Like the AD test, the Dip test is also a
non-parametric CDF statistic.  Where they differ is that the Dip test
looks for a flattening of the CDF for the data which would correspond
to a `dip' in the distribution being tested.  The Dip test operates
under the null hypothesis that the data is unimodal, and we consider a
group velocity distribution unimodal if the Dip test p-value is
greater than or equal to 0.05.  Therefore our G data sample consists
of all those groups with $p_{\text{ad}} \ge 0.05$
\emph{and} $p_{\text{dip}} \ge 0.05$, whereas our non-Gaussian (NG) data
sample consists of all those groups with $p_{\text{ad}} < 0.05$
\emph{or} $p_{\text{dip}} < 0.05$.
\par
After applying the above criteria we find a G sample consisting of
$42\,655$ galaxies within $2\,447$ groups and a NG sample consisting of $5\,306$
galaxies within 215 groups.  We find that the AD test is the stronger discrimator
compared to the Dip test as out of all of the galaxies making up the
NG sample, 90 per cent failed the AD test but passed the Dip test, 8
per cent passed the AD test but failed the Dip test, and 2 per cent
failed both the AD test and the Dip test.  The authors note that it is
easier to statistically identify NG groups for groups with high
galaxy membership, this can lead to the NG sample being skewed toward
large halo masses (see Fig.~\ref{fig:mhdist_um}).  To address this we
match our G and NG samples by halo mass (as well as stellar mass and
redshift), as described in the following section.

\subsection{Matched data set}
\label{sec:match}

\begin{figure*}
  \centering
  \includegraphics[width=\textwidth]{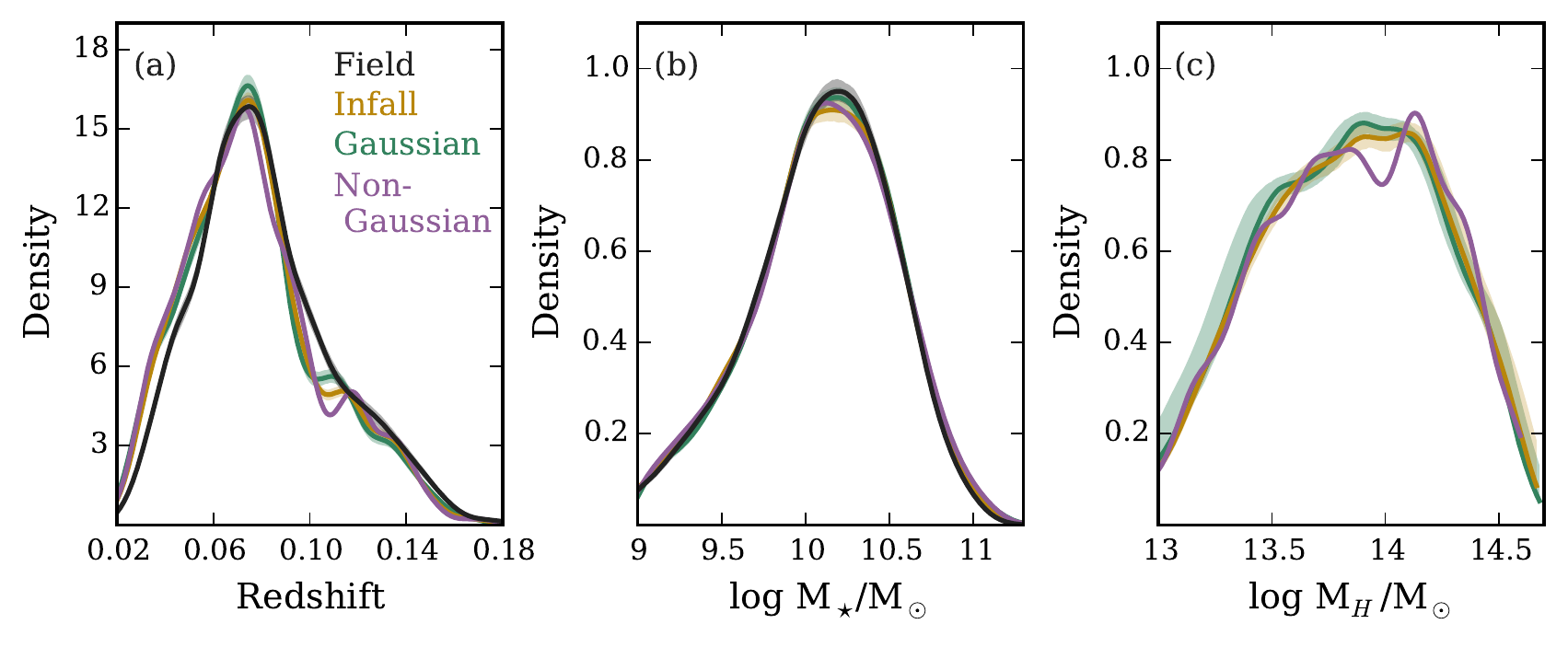}
  \caption{Distributions for stellar mass, redshift, and host
    halo mass for galaxies in the matched G, NG, infall, and field (where
    applicable) samples, smoothed using a Gaussian kernel.  Shaded regions around the G, infall, and field lines
    are 99 per cent Monte Carlo confidence intervals corresponding to the
    stochastic nature of our matching procedure.  The lines corresponding to the NG sample have no shading because it is the NG sample to which the other samples are stochastically matched.}
  \label{fig:dist_m2_s}
\end{figure*}

To ensure a fair comparison between galaxies in different environments
(ie.\ field galaxies, infall galaxies, galaxies in G groups, and galaxies in NG groups)
we match our sample of G group galaxies, NG group galaxies, and infalling galaxies by
stellar mass, redshift, and halo mass\footnote{Though galaxies in the infall sample are not identified as group members, we match them by the halo mass of the group they are nearest to in projection}.  Additionally, we then match
our sample of field galaxies by stellar mass and redshift.  The
matching is particularly important when trying to elucidate
information on the effect of group dynamics on galaxy star formation and
morphological properties for
two main reasons:
\par
First, stellar mass, redshift, and halo mass have
all been shown to influence galaxy star formation and morphology
\citep[e.g.][]{brinchmann2004, feulner2005, zheng2007, cucciati2012,
  wetzel2012, lackner2013, tasca2014}; whereas
the impact of group dynamics is less clear \citep{hou2013,
  ribeiro2013} which is suggestive of a more modest role.
Therefore,
to search for trends in galaxy star formation and morphology with group
dynamics it is crucial to properly control for these other known correlations.
\par
Second, standard statistical normality tests, such as the AD test, are
biased towards identifying non-Gaussian distributions when
sample size is large.  This is a result of the statistical power of
the test increasing with sample size which subsequently allows the
detection of more and more subtle departures from normality
\citep{razali2011}.  While these subtle departures from normality will
perhaps be statistically significant, they may not be physically relevant (in
principle, no group is perfectly Gaussian) and what really matters is
whether galaxies in groups which show large departures from normality
have different properties than galaxies in groups which show smaller
departures from normality. Since group
richness generally scales with halo mass, in the absence of any matching
procedure, a sample of NG groups will be biased towards large halo
masses compared to a similar sample of G groups -- even though many
high halo mass NG groups may have been identified on the basis of very
small departures from normality.  Ensuring that our G and NG
samples have similar halo mass distributions allows us to make a
fairer comparison between the two samples.
\par
Our algorithm for matching the G and NG samples is as follows:

\begin{enumerate}
  \item Our list of galaxies found in NG groups is iterated through,
    for each galaxy one `matching' galaxy from the G sample is
    found.  To be considered matching the two galaxies must have
    stellar masses within $0.1\,\mathrm{dex}$, redshifts within 0.01,
    and halo masses within $0.1\,\mathrm{dex}$.

  \item Step 1 is repeated until no more matches are
    found.  The end result is a list of galaxies from the NG
    sample each of which will have one or more matching galaxies from
    the G sample assigned to them
  
  \item The matched G sample is generated by including two galaxies
    from the G sample for every one matching galaxy from the NG
    sample.  By definition this excludes any galaxies in the NG
    sample which only have only one identified match.  However, 85 per cent
    of galaxies
    in the NG sample have two or more matches so although we reduce
    the NG sample size by 15 per cent it allows us to increase the
    matched G sample size twofold.  It is worth noting that when we
    run our analysis keeping only one matched G galaxy instead of two,
    we find no changes in the trends observed.

  \item In the case where a given galaxy in the NG sample has more
    than two identified matches, the two matching galaxies from the G
    sample are
    chosen randomly.  This introduces a stochastic nature to our
    analysis as each generation of the matched G sample will not
    contain exactly the same galaxies.  To account for this, any
    quantities calculated using the matched G sample are done so in a
    Monte Carlo sense where the median of 1000 stochastic generations
    is quoted.
\end{enumerate}

\noindent
The infall and field sample are subsequently matched to the NG sample following
the same procedure and the same method is used to account for the
stochastic nature of the matching procedure.  Fig.~\ref{fig:dist_m2_s}
shows smoothed density
distributions of stellar mass, redshift, and halo mass for the matched
G, NG, infall, and field samples.  For the remainder of the
paper all analysis is done using the matched samples, therefore from
this point forward any reference to the G, NG, infall, or field samples refers to the matched samples.


\begin{figure*}
  \centering
  \includegraphics[width=\textwidth]{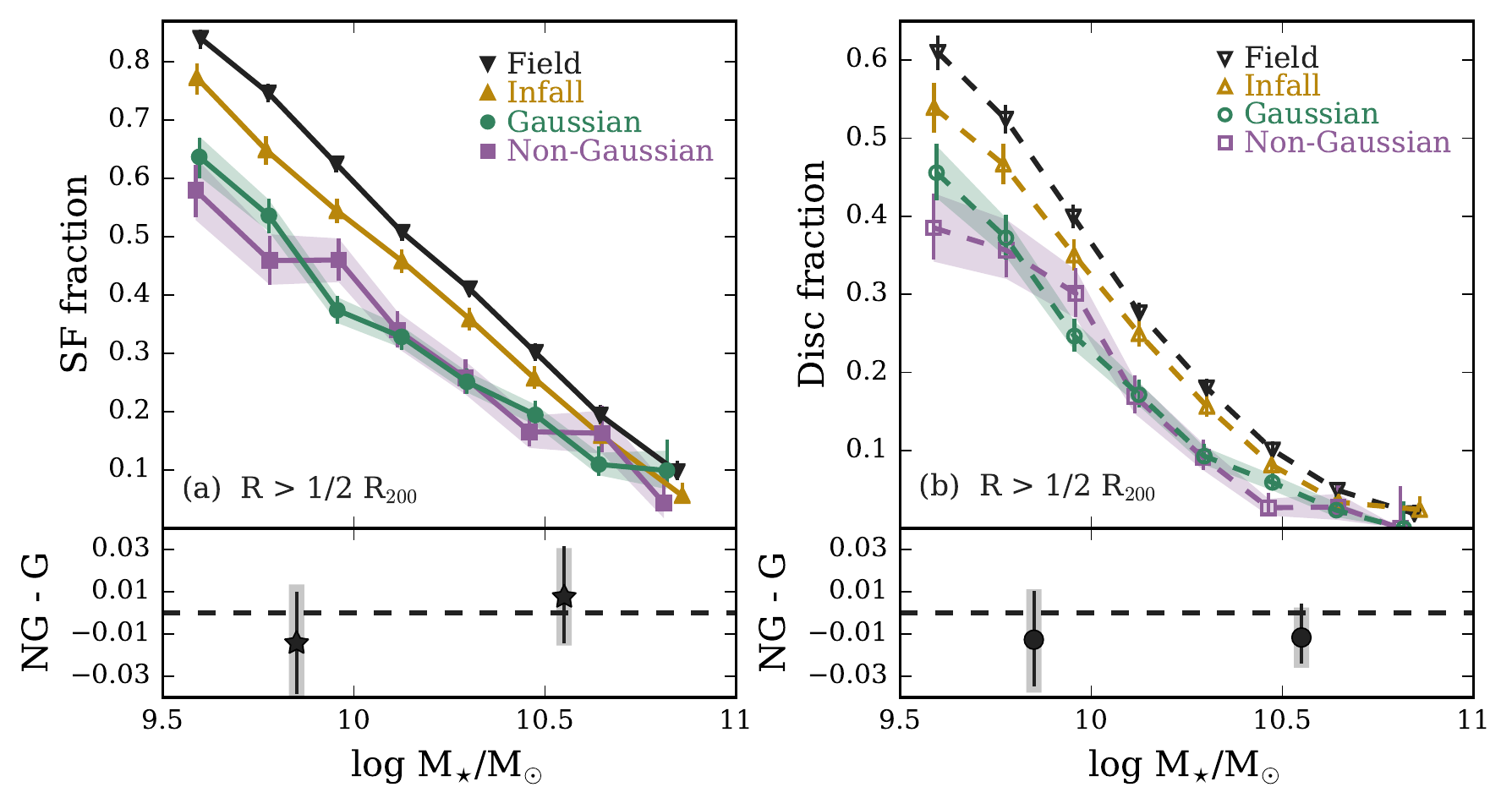}
  \caption{Star-forming (left) and disc (right) fraction versus stellar mass for
    field galaxies, infalling galaxies, and galaxies at large radii (outside $1/2\,\mathrm{R_{200}}$) in the G and NG
    samples.  Error bars correspond to 68 per cent binomial confidence
    intervals as given in \citet{cameron2011}, shaded regions are 68 per cent confidence intervals derived from 1000 bootstrap re-samplings over individual groups.  Lower panels show the difference in star-forming/disc fractions between G and NG groups, for low-mass ($M_\star \la 10^{10.2}\Msun$) and high-mass ($M_\star \ga 10^{10.2}\Msun$) galaxies.}
  \label{fig:disk_sfFrac_rmed1}
\end{figure*}

\begin{figure*}
  \centering
  \includegraphics[width=\textwidth]{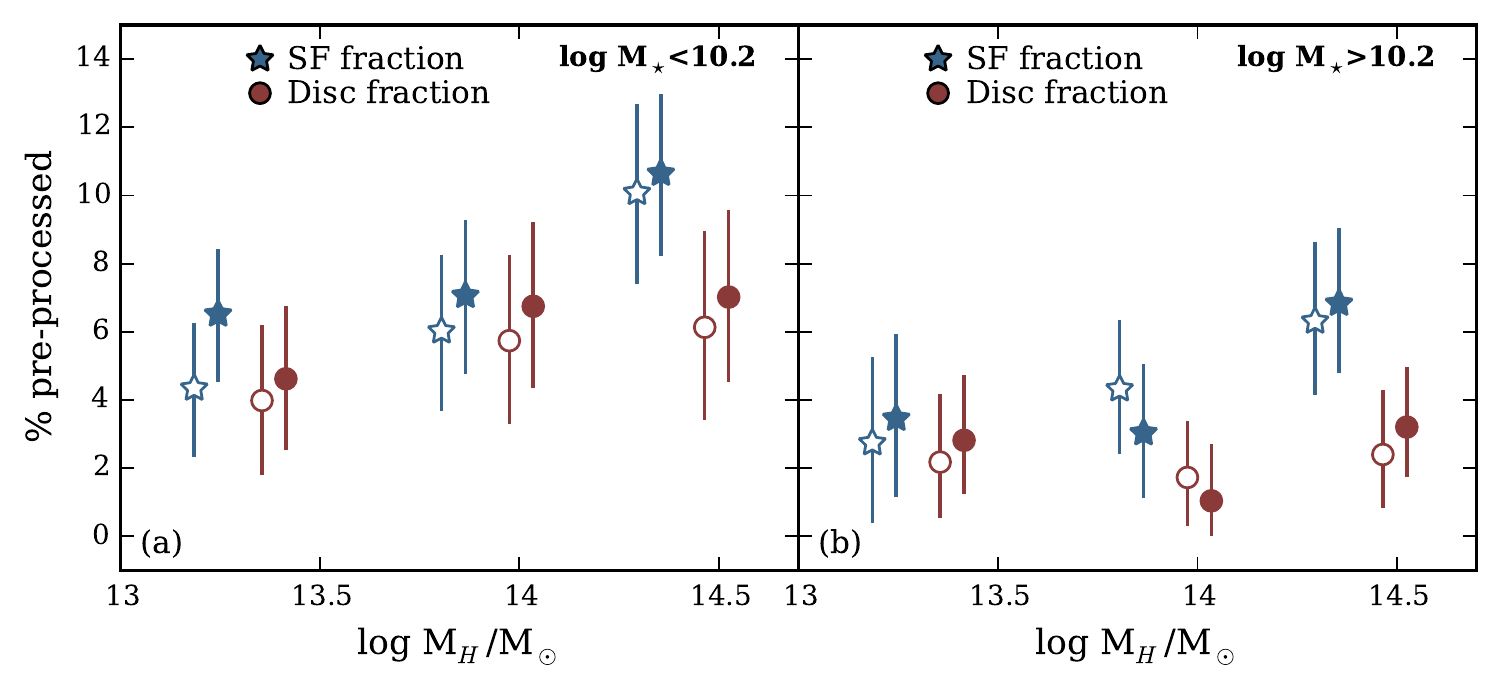}
  \caption{Pecentage of infalling galaxies which have had
    star formation (stars) or morphology (circles) pre-processed for
    both low-mass ($M_\star < 10^{10.2}\Msun$, left) and 
  high-mass ($M_\star > 10^{10.2}\Msun$, right) galaxies, as
  a function of halo mass.  Filled markers correspond to the whole infall sample ($1 < R < 3\,\mathrm{R_{200}}$) and open markers correspond to the strict infall sample ($2 < R < 3\,\mathrm{R_{200}}$).  Error bars are 68 per cent binomial confidence intervals \citep{cameron2011}.}
  \label{fig:pp_mh}
\end{figure*}

\section{Galaxy properties at large radii}
\label{sec:r_high}

We first consider the star-forming and morphological properties of
galaxies at large group-centric radii, and for comparison show the same trends for galaxies within the infall and field samples.  We separate galaxies at large and small radii at $1/2\,\mathrm{R_{200}}$ which is close to the median group-centric radius for the sample of $0.43\,\mathrm{R_{200}}$.  We apply
a lower stellar mass cut at $10^{9.5}\Msun$ in order to avoid
including galaxies with large $1/V_\mathrm{max}$ weights.
\par
In Fig.~\ref{fig:disk_sfFrac_rmed1} we show star-forming ($\log \mathrm{SSFR} > -11$) and disc ($n < 1.5$) fractions
versus stellar mass for the four different galaxy samples.  The bottom panels show the difference in star-forming/disc fractions between the NG and G samples (NG -- G) coarsely binned into low-mass ($M_\star < 10^{10.2}\Msun$) and high-mass ($M_\star > 10^{10.2}\Msun$) galaxies, where $10^{10.2}\Msun$ is the median stellar mass of the sample.  We estimate uncertainties on star-forming and disc fractions using two methods.  First, we follow \citet{cameron2011} who advocate the use of Bayesian binomial confidence intervals derived from the quantiles of the beta distribution to estimate statistical uncertainties on population fractions.  The error bars on the fractions correspond to 68 per cent confidence intervals obtained using this method.  Second, we quote 68 per cent bootstrap confidence intervals derived by bootstrapping over the member galaxies of individual groups, the confidence intervals derived from 1000 bootstrap realizations are shown as shaded regions.
\par
In Fig.~\ref{fig:disk_sfFrac_rmed1} we see a distinct trend in terms of star-forming and disc fractions, where field galaxies show the highest fractions, followed by infalling galaxies, followed by large-radius group members. Focusing now on the two dynamical samples we see no systematic difference
between the star-forming or disc fractions for galaxies at large-radii within
G groups compared to galaxies in NG groups.  This suggests that any influence that the dynamical state of the group
has on star-forming or morphological properties is not in place at large radii within the groups.  This is apparent in the lower panels of Fig.~\ref{fig:disk_sfFrac_rmed1} where the value of NG -- G is consistent with zero for both star-forming and disc fractions, regardless of stellar mass.  As stated in Section~\ref{sec:grp_dyn} we have used a p-value of 0.05 to divide the sample into G and NG groups, however we note that the results in Fig.~\ref{fig:disk_sfFrac_rmed1} are not sensitive to the specific choice from a reasonable range of p-values (see Appendix ~\ref{sec:appendix}).
\par
We also see that the star-forming and disc fractions for galaxies at large radii are significantly below the values for the field sample.  Previous studies \citep{lewis2002, gray2004, rines2005, verdugo2008}
have similarly found that star formation of galaxies within infall regions
remains suppressed compared to the field out to radii of
$\sim\!2-3\,R_{200}$.  This suppression is often attributed to
backsplash galaxies which have already made a passage through the halo
centre, the pre-processing of galaxies in small groups prior to
infall, or some combination of the two.  We are particularly interested in determining how much of this difference can be accounted for by pre-processing.
It is expected that pre-processing
should play a more important role in large clusters compared to smaller
groups, as a larger fraction of galaxies infalling onto clusters will
have been a part of a group prior to infall.  This is
a result of the hierarchical build-up of structure; regions of space around large clusters are not average but
are preferentially populated with other dense structures such as group
haloes \citep[e.g.][]{mo1996, wang2008}.
\par
We look for evidence of pre-processing by examining the ``field
excess'', which we define as the difference in star forming or disc
fraction between field and infalling galaxies at a given stellar mass,
for different halo mass ranges.  The range in group-centric radii for galaxies in the infall sample ($1 < R < 3\,\mathrm{R_{200}}$) is susceptible to contamination from galaxies backsplashing beyond the virial radius after first pericentric passage \citep[e.g.][]{bahe2013}.  To address this, we also show pre-processing results for our `strict' infall sample ($2 < R < 3\,\mathrm{R_{200}}$) which should be less susceptible to backsplash contamination, as many previous studies have shown that the majority of backsplashing galaxies are found within two virial radii \citep{mamon2004, mahajan2011, oman2013, haines2015}.  If contamination from backsplash galaxies is low, this field excess should approximate the fraction of galaxies which have been pre-processed prior to infalling onto their present-day group.  We investigate the halo mass dependence of pre-processing by splitting the group
sample into three halo mass bins each containing an approximately equal number of galaxies: $10^{13} < M_H \le 10^{13.7}\Msun$,
  $10^{13.7} < M_H \le 10^{14.1}\Msun$, and $10^{14.1} < M_H \le 10^{15}\Msun$, as well as two stellar mass bins (for each range in halo mass): $M_\star < 10^{10.2}\Msun$ and $M_\star > 10^{10.2}\Msun$.  In Fig.~\ref{fig:pp_mh} we show
the percentage of low-mass and high-mass galaxies which have been pre-processed in terms of star-forming fraction and disc fraction, and its dependence on halo mass, for the infall sample (filled markers) and the strict infall sample (open markers).
\par
For low-mass galaxies we detect modest, but statistically significant, pre-processing of both star formation and morphology (Fig.~\ref{fig:pp_mh}a).  At low stellar mass, the pre-processing of star formation tends to increase with halo mass, while the pre-processing of morphology shows a weaker halo mass trend.  Pre-processing trends for high-mass galaxies (both in terms of the magnitude of pre-processing, and halo mass trends) are much weaker.


\section{Galaxy properties at small radii}
\label{sec:r_low}

\begin{figure*}
  \centering
  \includegraphics[width=\textwidth]{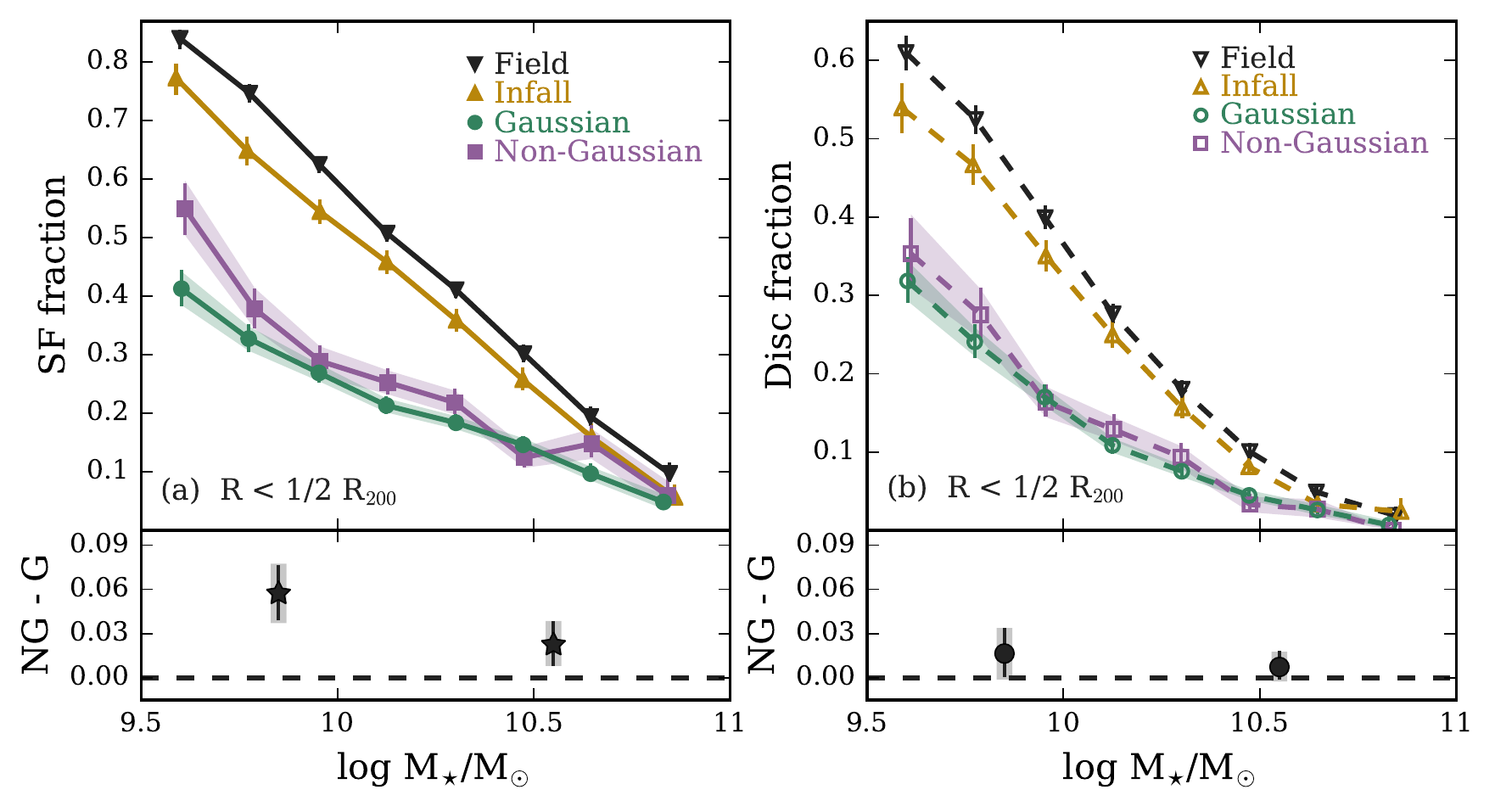}
  \caption{Star-forming (left) and disc (right) fraction versus stellar mass for
    field galaxies, infalling galaxies, and galaxies at small radii (within the median group-centric radius) in the G and NG
    samples.  Error bars correspond to 68 per cent binomial confidence
    intervals as given in \citet{cameron2011}, shaded regions are 68 per cent confidence intervals derived from 1000 bootstrap re-samplings over individual groups.  Lower panels show the difference in star-forming/disc fractions between G and NG groups, for low-mass ($M_\star \la 10^{10.2}\Msun$) and high-mass ($M_\star \ga 10^{10.2}\Msun$) galaxies.}
  \label{fig:disk_sfFrac_rmed0}
\end{figure*}

We now consider star-forming and disc fractions for galaxies at small radii within the halo, and again consider the differences between
the G, NG, infall, and field samples.  Fig.~\ref{fig:disk_sfFrac_rmed0} shows
star-forming and disc fractions versus stellar mass for the four
galaxy samples.  In contrast to the outer region of the halo, when
considering star-forming fractions for galaxies at small radii a dependence 
on group dynamics emerges.  In particular, galaxies in G groups have
the lowest star-forming fractions and galaxies in NG groups have intermediate
values -- larger star-forming fractions than galaxies in G
groups but significantly smaller than infalling galaxies or the field.  As shown in the lower panels of Fig.~\ref{fig:disk_sfFrac_rmed0}, this difference between G and NG groups is significant for low mass galaxies ($2.9\sigma$) but not for high mass galaxies ($1.6\sigma$).  When considering disc fraction, we do not detect a significant enhancement in NG groups for low or high mass galaxies ($0.9\sigma$ and $0.8\sigma$, respectively).  As was the case for galaxies at large radii, we show results corresponding to different p-value choices in Appendix~\ref{sec:appendix}.  The observed trends do not depend strongly on p-value.  If anything, choosing a larger p-value only strengthens the observed difference between G and NG at large stellar mass (see Appendix~\ref{sec:appendix}). 


\section{Discussion}
\label{sec:discussion}

\subsection{The impact of group dynamical state}

The question of how much group dynamical state influences galaxy
properties has not yet been conclusively answered.  In
this study we find that star formation of galaxies within the inner
regions of haloes show a dependence on group dynamics.  In
particular, we find that compared to G groups galaxies at small radii in NG
groups show an increase in star-forming fraction.
\par
\citet{carollo2013} study the differences between galaxies in
`relaxed' and `unrelaxed' groups (defined based upon the presence, or
lack thereof, of a well defined central group galaxy) in the Zurich
Environmental Study.  \citet{carollo2013} find that $<10^{10}\Msun$
satellites show slightly redder colours in relaxed groups compared to
unrelaxed groups.  Given the general correlations between galaxy colour and star
formation, this agrees well with the findings of this
work. \citet{ribeiro2013} use a statistical metric designed to quantify the
distance between probability density functions, known as the Hellinger
distance, to discriminate between G and NG groups using a
FOF catalogue of SDSS group galaxies
\citep{berlind2006}.  They find no dependence on group dynamics for
bright galaxies ($\mathrm{M_r} \le -20.7$), however find that
properties of faint galaxies ($-20.7 < \mathrm{M_r} \le -17.9$) do
depend on whether they live in a G or NG group.  Relevant to this
work, \citet{ribeiro2013} show that faint galaxies in G groups are
redder than their NG counterparts. As well, \citet{ribeiro2010} find
that galaxies in G groups are redder than galaxies in NG groups out to $4\,\mathrm{R_{200}}$.
\par
\citet{hou2013} have explored the dependence of quiescent fraction on group dynamical state as a function of redshift using a combination of groups from the SDSS and the Group Environment and Evolution Collaboration (GEEC).  For their low redshift galaxies, Hou et al. find no difference between the quiscent fraction of galaxies in NG versus G groups, though they use a stellar mass complete sample and are only able to probe masses of $M_\star > 10^{10}\Msun$.  This is consistent with the result from this work showing that any correlations with dynamical state are subtle and only seen for low-mass galaxies.
\par
The results of this paper can be used to further constrain the
connection between group dynamics and the quenching of star formation as well
as morphological transformations.  The main result is that we observe a
dependence of star formation on dynamics in the in the inner region of the halo but not
for galaxies at large radii, whereas morphology is not found to correlate with dynamics at any radius.  This seems to suggest that
quenching is primarly taking
place near the centres of groups, and is more efficient in G
groups than NG groups.  Alternatively, the observed excess of
star-forming galaxies in NG groups could be due to the more
dynamically complex NG groups having assembled more recently,
therefore galaxies in G groups will have been exposed to quenching
mechanisms within the group environment for longer.
\par
It is also worth noting that the unusual structure in velocity space of the NG groups could be a result of poorly identified groups which have undergone `fusing' (ie. two separate haloes which group finders have combined into one group) or `fracturing' (ie. one distinct halo which has been split into multiple groups by group finders).  Recent works \citep{duarte2014, campbell2015} have investigated the degree to which standard group finding techniques can accurately reproduce groups from mock catalogues.  \citet{campbell2015} show that these misidentifications can bias some colour-dependent statistics, such as red fraction which is directly related to the star-forming fraction considered in this work.  It would be useful in future work to apply the same statistics used here to discriminate between G and NG groups on mock catalogues in order to determine what fraction of identified NG groups are in fact unrelaxed, dynamically young systems as opposed to systems which have simply been misidentified by the group finder.

\subsection{Pre-processing of infalling galaxies}

In addition to star formation quenching and morphological
transformations within the current host halo, we also find evidence
for pre-processing in both star formation and morphology.  To probe
pre-processing we measure the ``field excess'' (ie.\ the degree to
which star-forming and disc fractions are enhanced in the field
relative to the infalling region of groups).  Assuming that any
environmentally driven quenching or morphological transformations occur within the virial radius of a halo,
the field excess will correspond to the fraction of
infalling galaxies which have been pre-processed.  Using this we
quantitatively determine the level of pre-processing by computing the
field excess for low-mass and high-mass
galaxies (divided at the median stellar mass of our sample, $M_\star \ga 10^{10.2}\Msun$) in our three halo mass bins.  As shown in Fig.~\ref{fig:pp_mh}, we find that the fraction of pre-processed low-mass galaxies ranges between 4 and 11 per cent when considering star-forming fraction and between 4 and 7 per cent when considering disc fraction.  For high-mass galaxies the pre-processed fraction is smaller and generally only marginally significant.
\par
Prior studies have aimed to constrain the
fraction of pre-processed galaxies.  One common approach is to measure
the fraction of
galaxies which fall onto a cluster as a member of a smaller group,
either directly using simulations or by measuring substructure or
clustering observationally.  For clusters with mass $\sim
\!10^{14}\Msun$ \citet{delucia2012} use semi-analytic models (SAMs) and find
that the fraction of satellite galaxies which are accreted in groups
with $M_H \ga 10^{13}\Msun$ is highest for low-mass galaxies,
corresponding to $\sim\!28$ per cent.  Also using SAMs,
\citet{mcgee2009} find that the fraction of galaxies accreted onto the
ultimate cluster as members of $\ga 10^{13}\,h^{-1}\Msun$ groups depends
strongly on the cluster halo mass, ranging from $\sim\!0.1$ for
$10^{13.5}\,h^{-1}\Msun$ haloes to $\sim\!0.45$ for haloes with masses
of $10^{15}\,h^{-1}\Msun$.  \citet{bahe2013} use the \textsc{gimic}
suite of zoom-in simulations and find that the fraction of galaxies
which have been satellites of a $>\!10^{13}\Msun$ halo prior to accretion
onto the ultimate host ranges from $<10$ per cent for a host with halo
mass $<\!10^{13.5}\Msun$, up to as high as $\sim\!60$ per cent for a
host halo mass of $10^{15.2}\Msun$.
Observationally, \citet{hou2014} use the
Dressler-Schectman test \citep{dressler1988} to identify infalling
subhaloes and find for $<\!10^{14}\Msun$ groups that less than 5 per
cent of infalling galaxies are part of a subhalo, whereas for haloes
with masses $10^{14} < M_H < 10^{14.5}\Msun$ and $M_H >
10^{14.5}\Msun$ the fraction of galaxies infalling in subhaloes is
$\sim 10$ per cent and $\sim 25$ per cent, respectively.
Qualitatively the pre-processing trends observed in this work are
consistent with these previous studies, namely the fraction of
pre-processed galaxies tends to decrease with increasing galaxy stellar mass and
increase with the halo mass of the host which the galaxies are
infalling onto.  The subhalo fraction found in these works can be interpretted as an upper limit on the field excess quantity which we
quote.  This is because only some fraction of galaxies within subhaloes
during infall will be pre-processed, whereas the
field excess more closely measures the fraction of galaxies
which have actually been pre-processed.  Therefore the fact that our
values for the fraction of pre-processed galaxies are consistently
smaller than the quoted subhalo fractions is still consistent.
\par
Studying the star-forming fractions of cluster galaxies,
\citet{haines2015} use a simple toy model in an attempt to reproduce the
trend between cluster-centric radius and star-forming fraction.  They
find that in order to reproduce the observational trend, a $19$ per
cent decrease in the star-forming fraction of cluster galaxies relative to
the field is required on top of star
formation quenching occuring within the virial radius.  Haines et
al. suggest that pre-processing is a possible mechanism to generate
this $19$ per cent decrease.  In this work we find that the fraction of
high-mass (the Haines et al. sample consists of galaxy stellar masses
$>\!2 \times 10^{10.2}\Msun$) pre-processed galaxies for high-mass
clusters is at most $7 \pm 2$ per cent.  Therefore, this work is only able to
account for a portion of the amount of pre-processing required by the
\citet{haines2015} model, although a more complete comparison would
require samples matched in halo mass and galaxy stellar mass.
\par
At $\sim\!z=0.2$, \citet{lu2012} find that blue fractions of low and
intermediate-mass cluster galaxies are lower than the field values
(at the same stellar mass) out to radii of $7\,\mathrm{Mpc}$, however
the most massive galaxies
show no difference from the field.  This is similar to the stellar
mass trends observed in this work where we see stronger
pre-processing for low-mass galaxies.
\par
Recent studies have examined pre-processing of morphology
\citep[e.g.][]{kodama2001, helsdon2003, moran2007, wilman2009} and star
formation \citep[e.g.][]{cortese2006, wetzel2012, bahe2013,
  haines2015} separately, however we are not aware of other works which have made direct quantitative comparisons between the amount of pre-processing of star formation versus morphology.  In Fig.~\ref{fig:pp_mh} we see evidence for pre-processing in both star formation and morphology, though due to the relatively large errorbars it is unclear whether one is more strongly pre-processed than the other.  The largest difference between star formation and morphology is in the highest halo mass bin, where star formation shows marginally stronger pre-processing than morphology.  Additionally if we consider the entire data set (without subdividing by halo mass) we find that the pre-processing of star formation rate is marginally enhanced relative to morphology at the $\sim\! 2\sigma$ level.  Understanding the relative strength of pre-processing of star formation versus morphology could help to disentangle environmentally driven galaxy evolution mechanisms and should be explored further.


\section{Summary \& conclusions}
\label{sec:summary}

In this paper we investigate the dependence of galaxy properties
(namely, star-forming and disc fractions) on host group dynamics.  To
do so we construct a carefully matched sample of galaxies housed in
Gaussian groups, galaxies housed in non-Gaussian groups, as well as
infalling and field galaxies; all with similar distributions in stellar mass,
redshift, and (field galaxies excluded) halo mass.  We then compare
the properties of these different samples for two different radial regions within the halo.  The main findings of this work are as follows:

\begin{enumerate}
  \item Star-forming and disc fractions of galaxies at large group-centric radius do not show any dependence on the dynamical state of their host group

  \item We detect pre-processing by measuring the difference between
    the star-forming and disc fractions for field galaxies compared to
    infalling galaxies.  Infalling galaxies have had both star
    formation and morphology pre-processed, with low-mass galaxies
    infalling onto high-mass haloes showing the largest degree of
    pre-processing.

  \item Galaxy star-formation in the inner region of the halo shows a clear dependence on group dynamical state, with enhanced star-forming fractions for galaxies in non-Gaussian groups compared to galaxies in Gaussian groups at the same stellar mass.  We do not detect a significant dependence of disc fraction on group dynamical state in the same inner region.
\end{enumerate}


\section*{Acknowledgments}
\label{sec:acknowledgments}

We thank the referee for their insightful comments and suggestions, which have improved this paper significantly.  IDR thanks the Ontario Graduate Scholarship program and the National Science and Engineering Research Council of Canada for funding.  LCP
thanks the National Science and Engineering Research
Council of Canada for funding.  The authors thank
F. Evans for matching together the various SDSS catalogues used in
this research.  We thank X. Yang et al. for
making their
SDSS DR7 group catalogue publicly available, L. Simard et al. for the
publication of their SDSS DR7 morphology catalogue, J. Brinchmann et al. for
publication of their SDSS SFRs, and the NYU-VAGC
team for the 
publication of their SDSS DR7 catalogue.  This research would not have
been possible without access to these public catalogues.
\par
Funding for the SDSS has been provided by the Alfred P. Sloan
Foundation, the Participating Institutions, the National Science
Foundation, the U.S. Department of Energy, the National Aeronautics
and Space Administration, the Japanese Monbukagakusho, the Max Planck
Society, and the Higher Education Funding Council for England. The
SDSS Web Site is http://www.sdss.org/.
\par
The SDSS is managed by the Astrophysical Research Consortium for the
Participating Institutions. The Participating Institutions are the
American Museum of Natural History, Astrophysical Institute Potsdam,
University of Basel, University of Cambridge, Case Western Reserve
University, University of Chicago, Drexel University, Fermilab, the
Institute for Advanced Study, the Japan Participation Group, Johns
Hopkins University, the Joint Institute for Nuclear Astrophysics, the
Kavli Institute for Particle Astrophysics and Cosmology, the Korean
Scientist Group, the Chinese Academy of Sciences (LAMOST), Los Alamos
National Laboratory, the Max-Planck-Institute for Astronomy (MPIA),
the Max-Planck-Institute for Astrophysics (MPA), New Mexico State
University, Ohio State University, University of Pittsburgh,
University of Portsmouth, Princeton University, the United States
Naval Observatory, and the University of Washington.



\bibliographystyle{mnras}
\bibliography{CompleteManuscriptFile_accepted} 


\appendix

\section{Dependence on the definition of NG groups}
\label{sec:appendix}

\begin{figure*}
  \centering
  \includegraphics[width=\textwidth]{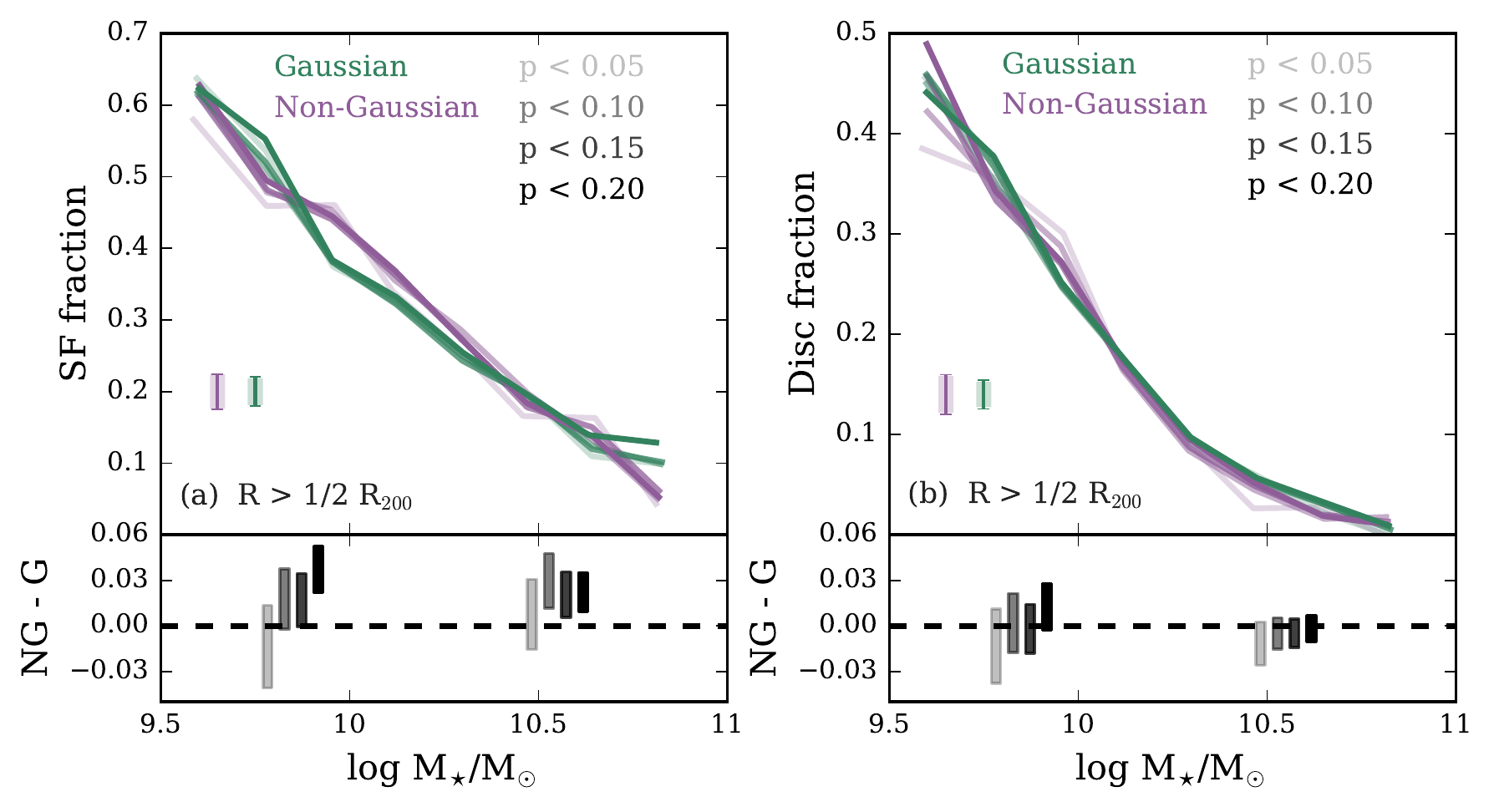}
  \caption{Star-forming (left) and disc (right) fraction versus stellar mass for
    galaxies at large radius in the G and NG samples.  The lines of varying transparency correspond to different definitions of the NG sample, where the listed p-value is the critical value used in the AD and Dip tests to identify NG groups.  Characteristic uncertainties are shown for 68 per cent confidence intervals from \citet{cameron2011} (errorbars) and from 1000 bootstrap re-samplings (shaded regions).  Lower panels show the difference between star-forming fractions in G and NG groups (left) and similarly for disc fraction (right), for low-mass ($M_\star \la 10^{10.2}\Msun$) and high-mass ($M_\star \ga 10^{10.2}\Msun$) galaxies.}
  \label{fig:disk_sfFrac_rmed1_p}
\end{figure*}

\begin{figure*}
  \centering
  \includegraphics[width=\textwidth]{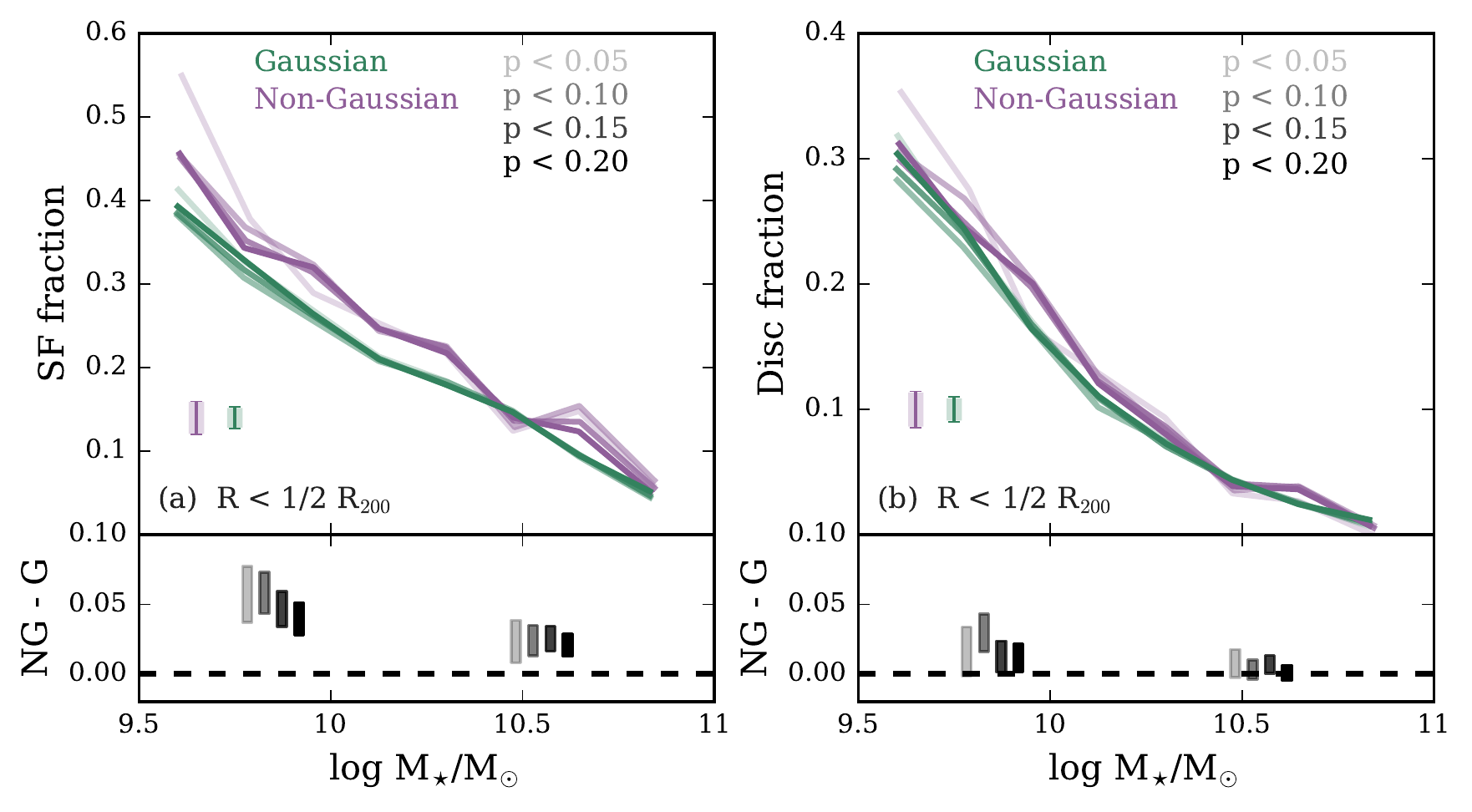}
  \caption{Star-forming (left) and disc (right) fraction versus stellar mass for
    galaxies at small radius in the G and NG samples.  The lines of varying transparency correspond to different definitions of the NG sample, where the listed p-value is the critical value used in the AD and Dip tests to identify NG groups.  Characteristic uncertainties are shown for 68 per cent confidence intervals from \citet{cameron2011} (errorbars) and from 1000 bootstrap re-samplings (shaded regions).  Lower panels show the difference between star-forming fractions in G and NG groups (left) and similarly for disc fraction (right), for low-mass ($M_\star \la 10^{10.2}\Msun$) and high-mass ($M_\star \ga 10^{10.2}\Msun$) galaxies.}
  \label{fig:disk_sfFrac_rmed0_p}
\end{figure*}

To discriminate between G and NG groups we use a critical p-value of 0.05 for both the AD test and the Dip test (see Section~\ref{sec:grp_dyn}).  While this choice of 0.05 is standard, it is still an arbitrary choice and it is therefore important to investigate the effect of varying this dividing p-value.
\par
Figs~\ref{fig:disk_sfFrac_rmed1_p} and \ref{fig:disk_sfFrac_rmed0_p} show star-forming and disc fractions for galaxies in the outer and inner regions of G and NG groups (similar to Figs~\ref{fig:disk_sfFrac_rmed1} and \ref{fig:disk_sfFrac_rmed0}), for different choices of the dividing p-value between G and NG groups.  The lower panels in Figs~\ref{fig:disk_sfFrac_rmed1} and \ref{fig:disk_sfFrac_rmed0} show NG -- G for different choices of p-value, with the height of the data marker corresponding to 68 per cent confidence intervals derived from either bootstrapping or the methodology of \citet{cameron2011} (whichever is larger).  Please note that the data markers are offset from one another for visibility.  Lines range in decreasing transparency from p-values of 0.05 to 0.20.
\par
We see no indication that the results of this work are strongly sensitive to the choice of p-value, as qualitatively similar trends are observed for all p-values shown in Figs~\ref{fig:disk_sfFrac_rmed1_p} and \ref{fig:disk_sfFrac_rmed0_p}.  Quantitatively, no enhancement of disc fractions in NG groups is seen above the $2\sigma$ level at large or small radius (and for large or small stellar mass), with the exception of the $p<0.10$ sample at small stellar mass where an enhancement of $2.1\sigma$ is detected.  Considering star-forming fractions at small radius we see a similar enhancement of star-forming fractions in NG groups for low-mass galaxies, significant at $>2\sigma$ for all p-values.  For high-mass galaxies, an enhancement at the $2\sigma$ level is detected for the $p<0.10$, $p<0.15$, and $p<0.20$ cases whereas the $2\sigma$ level was not reached for the $p<0.05$ sample used in the paper.  This is primarily due to uncertainties on the fractions in the NG sample becoming smaller as the dividing p-value is increased (leading to a larger NG sample).  For star-forming fractions at large radius an enhancement in NG groups is only detected at $>2\sigma$ in the $p<0.20$ case for low-mass galaxies, and no enhancement is detected in any case for high-mass galaxies.  These trends with star-forming and disc fractions are generally consistent with the trends presented in the body of the paper, with any differences only strengthening the conclusion that star formation is enhanced in NG groups relative to G groups.


\bsp	
\label{lastpage}
\end{document}